# An AsFem implementation for quasi-static brittle fracture phase field model

Yuanfeng Yu[1], Xiaoya Zheng[2],[*]

*1. School of Aeronautics, Northwestern Polytechnical University, Xi'an 710072, China*

*2. School of Astronautics, Northwestern Polytechnical University, Xi'an 710072, China*

**Abstract**

Phase field method has been widely used because of its excellent ability to simulate fracture problems. At present, the implementation process is mainly based on commercial software, and the operation process is relatively complex. In this paper, 2D and 3D phase field models are implemented in the open-source finite element software package AsFem. Compared with commercial software, it is simpler to realize phase field fracture model in AsFem. At the same time, a robust staggered scheme and an efficient monolithic scheme can be used to solve the displacement field and phase field sub-problem, and the transformation process is very easy. Several examples are tested to demonstrate the performance of the current implementation. The simulation results are in good agreement with the previous work, which shows the feasibility and effectiveness of implementing the phase field method in AsFem. In the future, more complex multi-field coupled phase field fracture problems can be solved with AsFem.

# 1. Introduction

Since the cracks in solids lead to their performance degradation and even structural damage[1,2], the study of their failure mechanisms is very important in engineering applications. In particular, when certain types of crack growth experiments are difficult or even impossible to perform, researchers must employ numerical methods to predict complex crack paths. Therefore, a large number of numerical methods are used to deal with the problem of cracking.

---

[*] E-mail: zhengxy_8042@nwpu.edu.cn

In the field of structural and mechanical research, numerical analysis methods can usually be divided into discrete and diffusion methods. For discrete method, such as virtual crack closure technique (VCCT)[3], extended finite element method (XFEM)[4]. When the VCCT simulates crack propagation, the cracks need to be pre-arranged, and the meshes near the cracks need to be subdivided or continuously re-divided. The XFEM describes the discontinuous term by increasing the enrichment term of displacement, but its convergence is poor. And it is necessary to know the possible failure mode of the material in advance, and define the corresponding cracking direction after damage[4]. Diffusion numerical methods such as peridynamics[5], phase field method[6], in these methods, the crack is diffused to describe the crack propagation process. For the peridynamics, the deformation coordination equation and force balance equation in solid mechanics are rewritten as integral equations, so as to describe the crack initiation, propagation, branching and other problems[7,8]. However, this method is difficult to directly correspond to the constitutive model of the material[9].

Nowadays the phase field method has been widely concerned by researchers because of its flexible implementation. A scalar phase field is used to represent discrete cracks[10], which does not describe the crack as a physical discontinuity, but rather smoothly transitions from an intact state to a fully fractured state, with the growth of the crack depend on the governing equation of the phase field. In order to make the crack phase field evolution satisfy the irreversible constraint, Miehe et al.[11] introduced the history field to solve the crack propagation problem under cyclic loading. After this work, the phase field method has been continuously improved and applied to different research fields. Such as brittle fracture problem[12,13], plastic fracture problem[14,15], dynamic fracture problem[16,17], cohesive fracture problem[18,19], fatigue problem[20,21]. At the same time, the scope of research materials is also expanding, such as rock-like materials[22,23], polymers[24,25], concrete materials[26,27], composite materials[28,29]. Compared with other methods, such as the XFEM, the interface element method, the phase field method has its unique advantages in studying the material failure process, which enables it to describe discontinuous cracks and simulate complex forms of crack.

In recent years, the phase field method has been implemented not only in commercial software, such as ABAQUS[30,31] and COMSOL[32,33], but also in some open source software packages, such as FEniCS[34,35], FEAP[36], Deal.II[37], and MOOSE[38] and others. These commercial software have excellent secondary development capabilities, which are convenient for users to solve complex mechanical problems. However, for commercial software, the purchase cost is relatively high due to software copyright issues, and the software does not currently have the function of directly implementing the phase field method, so secondary development is required. For phase field researchers, it is necessary to have a certain finite element theory and to understand the secondary development rules of software. Simultaneously, although the model can also be implemented on the open source software package, the relative implementation process is more complicated and requires a solid finite element analysis foundation, which is not friendly to beginners studying the phase field method. Therefore, the actual effect will be overrated. For example, the FEniCS software package is composed of different modules, and it takes a lot of time to learn its low-level implementation process. If one wants to customize and modify some modules, the implementation will be very difficult, so it will take more time to implement the phase field method.

In this paper, the phase field fracture model is implemented in AsFem (A Simple Finite Element Method)[39] in both monolithic scheme and staggered scheme. AsFem is a new open source finite element package in which solid mechanics problems, Cahn Hilliard diffusion problems, phase field fracture problems, linear elasticity problems can be implemented. The phase field model can be easily implemented in AsFem without much finite element theory, providing an efficient computing platform for phase field researchers. In AsFem, an implicit time integration scheme is used for simulation. Some examples for crack propagation are used to show the feasibility of the fracture modeling approach in AsFem.

The rest of the paper is as follows. In Section 2, the basic theory of the phase field model is reviewed. In Section 3, the process of implementing phase field method is presented in AsFem using staggered and monolithic scheme. Section 4 presents the simulation results of some examples. Finally, the main conclusions are summarized in the section 5.

## 2. Review of the phase field method
### 2.1. Fracture energy functional

For a crack $\Gamma$ in a continuous solid, a finite width can be used $l$ to diffuse the crack, so that the discrete crack can be simulated by some continuous function, as shown in Fig. 1. In this way, auxiliary variables can be introduced $d(x) \in [0,1]$ to describe such sharp cracks, which $d=0$ represents the intact state of the solid, and $d=1$ represents the fully fractured state of the solid.

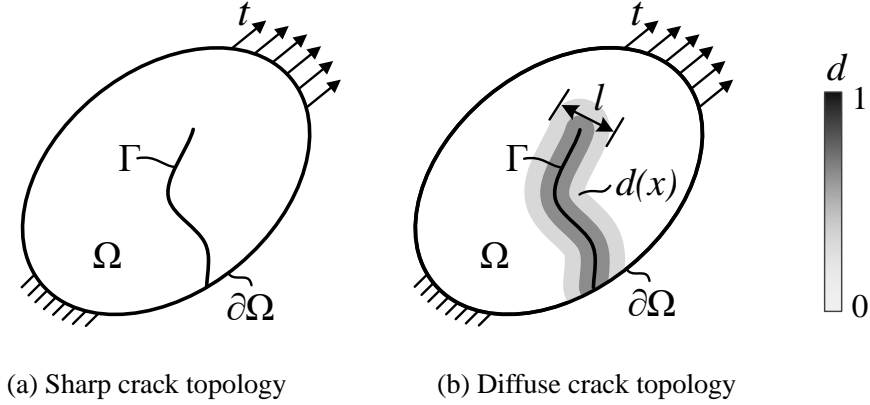

(a) Sharp crack topology    (b) Diffuse crack topology

Fig. 1. Crack topology

Therefore, the crack surface energy can be approximated by

$$\Gamma(d) = \frac{1}{2l} \int_\Omega \left\{ d^2 + l^2 |\nabla d|^2 \right\} dV \tag{1}$$

Here, the crack surface density function $\gamma$ and derivative per unit area (or length) $\delta_d \gamma$ are

$$\gamma(d, \nabla d) = \frac{1}{2l} d^2 + \frac{l}{2} |\nabla d|^2, \quad \delta_d \gamma = \frac{1}{l} d - l \Delta d. \tag{2}$$

For the phase field fracture problem in Fig. 1, according to the fracture surface function $\Gamma(d)$ introduced in Eq.(1), the work required to generate a new crack topology can be defined

$$\Pi_c(d) = \int_\Omega G_c \gamma(d, \nabla d) dV \tag{3}$$

Where $G_c$ is the Griffith-type critical energy release rate and is the work required to generate a new crack per unit length[40].

Meanwhile, according to the linear elasticity theory, the global energy storage functional can be expressed as

$$\Pi_e(\boldsymbol{u}, d) = \int_B \psi(\varepsilon(\boldsymbol{u}), d) dV, \tag{4}$$

For the energy reduction caused by fracture, it can be expressed as

$$\psi(\varepsilon,d) = [g(d)+k]\psi_0(\varepsilon). \tag{5}$$

In the Eq.(5), $g(d)$ is the energy degradation function, a simple function of $g(d)=(1-d)^2$. Parameter $k$ is the residual stiffness coefficient, to ensure numerical stability, $\psi_0$ is the strain energy function of the undamaged material, and for isotropic materials, it can be expressed as

$$\psi_0(\varepsilon) = \lambda \operatorname{tr}^2[\varepsilon]/2 + \mu \operatorname{tr}[\varepsilon^2]. \tag{6}$$

Among them, the Lame constant $\lambda > 0$ and $\mu > 0$. According to the spectral decomposition method[11], the strain energy is decomposed into positive part $\psi_0^+(\varepsilon)$ and negative part $\psi_0^-(\varepsilon)$, which can be expressed as

$$\begin{aligned}\psi_0^+(\varepsilon) &= \lambda \langle \operatorname{tr}(\varepsilon) \rangle_+^2 / 2 + \mu \operatorname{tr}(\varepsilon_+^2), \\ \psi_0^-(\varepsilon) &= \lambda \langle \operatorname{tr}(\varepsilon) \rangle_-^2 / 2 + \mu \operatorname{tr}(\varepsilon_-^2). \end{aligned} \tag{7}$$

## 2.2. Variational phase field model

According to the crack dissipation functional (4) and the energy storage functional equation (5), the total energy functional is introduced

$$\begin{aligned}\Pi(\boldsymbol{u},d) = &\int_B \psi(\varepsilon(\boldsymbol{u}),d)dV + \int_\Omega G_c \gamma(d,\nabla d)dV - \\ &\int_B \boldsymbol{b}\cdot\boldsymbol{u}dV - \int_{\partial B} \boldsymbol{t}\cdot\boldsymbol{u}dA.\end{aligned} \tag{8}$$

where is $b$ the volume force in the region $B$, and $t$ is the surface external force on $\partial B$.

Taking the variation of the total energy functional equation and considering the viscous effect, the following strong form governing equation is obtained

$$\begin{cases} \nabla\cdot\boldsymbol{\sigma} + b = 0 \\ f - G_c \delta_d \gamma - \eta \dot{d} = 0 \end{cases} \tag{9}$$

Here, the viscosity coefficient is taken as $\eta = 1\times 10^{-6}$, the energy driving force $f$ conjugated to the energy functional $\psi(\varepsilon,d)$ is

$$f := -\frac{\partial \psi}{\partial d} = -g'(d)\psi_0^+ \tag{10}$$

In order to ensure the irreversibility of damage evolution, a positive maximum historical reference energy is introduced

$$\mathcal{H}(\boldsymbol{x},t) := \max_{s\in[0,t]}\left[\psi_0^+\left(\varepsilon(\boldsymbol{x},t)\right)\right]. \tag{11}$$

Here, Substituting Eq.(11) for $\psi_0^+$ in Eq.(10), one can get

$$f := -g'(d)\mathcal{H}. \tag{12}$$

## 3. Implementation method in AsFem

### 3.1 Element discrete

For the Eq.(9), the weak form of the corresponding equation is expressed as

$$\begin{cases} \int_\Omega \boldsymbol{\sigma}:\delta\varepsilon dV - \int_\Omega \boldsymbol{b}\cdot\delta\boldsymbol{u}dV - \int_\Omega \boldsymbol{t}\cdot\delta\boldsymbol{u}dA = 0 \\ \int_\Omega g'(d)\delta d\mathcal{H} + \left(\dfrac{G_c}{l}d\delta d + G_c l \nabla d\cdot\nabla\delta d + \eta\dot{d}\delta d\right)dV = 0 \end{cases} \tag{13}$$

In Eq.(13), $\dot{d} := (d-d_n)/\tau$, $\tau$ is the time increment. Their corresponding allowable function and test function are shown in Eq.(14),

$$\begin{aligned} \mathbb{U}_u &:= H_1^E = \left\{\boldsymbol{u}\mid \boldsymbol{u}(x)=\boldsymbol{u}^*, \forall \boldsymbol{x}\in\partial\mathcal{B}_u\right\}, \\ \mathbb{V}_u &:= H_1^0 = \left\{\delta\boldsymbol{u}\mid \delta\boldsymbol{u}(x)=0, \forall \boldsymbol{x}\in\partial\mathcal{B}_u\right\}, \\ \mathbb{U}_d &:= H_1^E = \left\{d\mid d(x)\in[0,1], \dot{d}(x)\geq 0, \forall \boldsymbol{x}\in\Omega\right\}, \\ \mathbb{V}_d &:= H_1^0 = \left\{\delta d\mid \delta d(x)=0, \forall \boldsymbol{x}\in\Omega\right\}. \end{aligned} \tag{14}$$

The weak form Eq. (13) is usually discretized by multi field finite element method. Here, the displacement field $\boldsymbol{u}$ and its corresponding strain field $\varepsilon$ are first discretized

$$\boldsymbol{u}(x) = \sum_{i=1}^n N_i \boldsymbol{u}_i = \boldsymbol{N}\boldsymbol{u}, \quad \varepsilon(x) = \sum_{i=1}^n B_i \boldsymbol{u}_i = \boldsymbol{B}\boldsymbol{u}, \tag{15}$$

where, the interpolation matrix $\boldsymbol{N} := [\boldsymbol{N}_1,\cdots \boldsymbol{N}_i,\cdots,\boldsymbol{N}_n]$, and the geometric matrix $\boldsymbol{B} := [\boldsymbol{B}_1,\cdots,\boldsymbol{B}_i,\cdots,\boldsymbol{B}_n]$.

Similarly, the phase field $d$ and its corresponding gradient field $\nabla d$ can be discretized as

$$d(x) = \sum_{i=1}^n \bar{N}_i d_i = \bar{\boldsymbol{N}}\boldsymbol{d}, \quad \nabla d(x) = \sum_{i=1}^n \bar{B}_i d_i = \bar{\boldsymbol{B}}\boldsymbol{d}. \tag{16}$$

The corresponding heuristic function and its derivative are

$$\delta \boldsymbol{u}(x) = \sum_{i=1}^{n} \boldsymbol{N}_i \delta \boldsymbol{u}_i = \boldsymbol{N} \delta \boldsymbol{u}, \; \delta \varepsilon(x) = \sum_{i=1}^{n} \boldsymbol{B}_i \delta \boldsymbol{u}_i = \boldsymbol{B} \delta \boldsymbol{u}$$
$$\delta d(x) = \sum_{i=1}^{n} \bar{\boldsymbol{N}}_i \delta d_i = \bar{\boldsymbol{N}} \delta d, \; \nabla \delta d(x) = \sum_{i=1}^{n} \bar{\boldsymbol{B}}_i \delta d_i = \bar{\boldsymbol{B}} \delta d \quad (17)$$

After the above finite element discretization, the residual equation of displacement field and phase field can be expressed as

$$r^u = \int_\Omega (\boldsymbol{B})^\mathrm{T} \boldsymbol{\sigma} dV - \int_\Omega (\boldsymbol{N})^\mathrm{T} b dV - \int_{\partial\Omega} (\boldsymbol{N})^\mathrm{T} t dS = 0$$
$$r^d = \int_\Omega \left\{ \left[ \frac{G_c}{l} d + g'(d) \mathcal{H} + \eta \dot{d} \right] (\bar{\boldsymbol{N}})^\mathrm{T} + G_c l (\bar{\boldsymbol{B}})^\mathrm{T} \nabla d \right\} dV = 0 \quad (18)$$

## 3.2. Monolithic scheme

The monolithic scheme is to solve the displacement field and the phase field Simultaneously. Here, the above equation system (18) can be solved by the Newton-Raphson iteration method. According to the known displacement field $\boldsymbol{u}_n$, historical field $\mathcal{H}_n$ and phase field $d_n$ value at the time $t_n$, the calculation process is shown in the Fig. 2, this system of equations can be expressed in the following solution form

$$\begin{bmatrix} K^{uu} & K^{ud} \\ K^{du} & K^{dd} \end{bmatrix} \begin{bmatrix} \boldsymbol{u} \\ d \end{bmatrix} = -\begin{bmatrix} r^u \\ r^d \end{bmatrix} \quad (19)$$

where:

$$r^u = \int_\mathcal{B} (\boldsymbol{B})^\mathrm{T} \boldsymbol{\sigma} dV - \left( \int_\mathcal{B} (\boldsymbol{N})^\mathrm{T} b dV + \int_{\partial\mathcal{B}} (\boldsymbol{N})^\mathrm{T} t dA \right)$$
$$r^d = \int_\Omega \left\{ \left[ \frac{G_c}{l} d - 2(1-d) \mathcal{H} + \eta \dot{d} \right] (\bar{\boldsymbol{N}})^\mathrm{T} + G_c l (\bar{\boldsymbol{B}})^\mathrm{T} \nabla d \right\} dV \quad (20)$$

$$K^{uu} = \frac{\partial r^u}{\partial \boldsymbol{u}} = \int_\Omega (\boldsymbol{B})^\mathrm{T} \left[ \frac{\partial \boldsymbol{\sigma}}{\partial \varepsilon} \right] \boldsymbol{B} dV$$
$$K^{ud} = \frac{\partial r^u}{\partial d} = \int_\Omega (\boldsymbol{B})^\mathrm{T} \left[ \frac{\partial \boldsymbol{\sigma}}{\partial d} \right] \bar{\boldsymbol{N}} dV$$
$$K^{du} = \frac{\partial r^d}{\partial \boldsymbol{u}} = \int_\mathcal{B} -2(1-d)(\bar{\boldsymbol{N}})^\mathrm{T} \left[ \frac{\partial \mathcal{H}}{\partial \varepsilon} \right] \boldsymbol{B} dV$$
$$K^{dd} = \frac{\partial r^d}{\partial d} = \int_\mathcal{B} G_c l (\bar{\boldsymbol{B}})^\mathrm{T} \bar{\boldsymbol{B}} + \left[ \frac{G_c}{l} + 2\mathcal{H} + \frac{\eta}{\tau} \right] (\bar{\boldsymbol{N}})^\mathrm{T} \bar{\boldsymbol{N}} dV \quad (21)$$

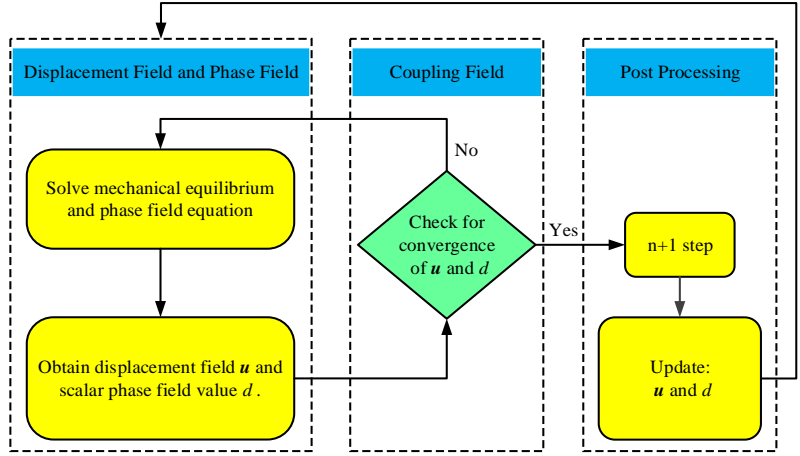

Fig. 2. Flowchart of the monolithic scheme

## 3.3. Staggered scheme

The staggered scheme is to solve the displacement field and the phase field independently, that is, use the phase field $d_n$ value at the moment $t_n$ to solve the displacement field $u_{n+1}$ at the moment $t_{n+1}$, then use the calculated displacement field value at moment $t_{n+1}$ to solve the phase field $d_{n+1}$ at the moment $t_{n+1}$, the calculation process is shown in the Fig. 3. This method is similar to a semi-implicit solution method, this system of equations can be expressed in the following solution form

$$\begin{bmatrix} K_n^d & 0 \\ 0 & K_n^u \end{bmatrix} \begin{bmatrix} d_{n+1} \\ u_{n+1} \end{bmatrix} = -\begin{bmatrix} r_n^d \\ r_n^u \end{bmatrix} \qquad (22)$$

where:

$$r_n^d = \int_\Omega \left\{ \left[ \frac{G_c}{l} d - 2(1-d)\mathcal{H} + \eta \dot{d} \right] (\bar{N})^{\mathrm{T}} + G_c l (\bar{B})^{\mathrm{T}} \nabla d \right\} dV$$

$$K_n^d = \frac{\partial r^d}{\partial d} = \int_\mathcal{B} \left\{ \left[ \frac{G_c}{l} + 2\mathcal{H} + \frac{\eta}{\tau} \right] (\bar{N})^{\mathrm{T}} \bar{N} + G_c l (\bar{B})^{\mathrm{T}} \bar{B} \right\} dV \qquad (23)$$

$$r_n^u = \int_\mathcal{B} (B)^{\mathrm{T}} \sigma dV - \left( \int_\mathcal{B} (N)^{\mathrm{T}} b dV + \int_{\partial \mathcal{B}} (N)^{\mathrm{T}} t dA \right)$$

$$K_n^u = \frac{\partial r^u}{\partial u} = \int_\Omega (B)^{\mathrm{T}} \left[ \frac{\partial \sigma}{\partial \varepsilon} \right] B dV \qquad (24)$$

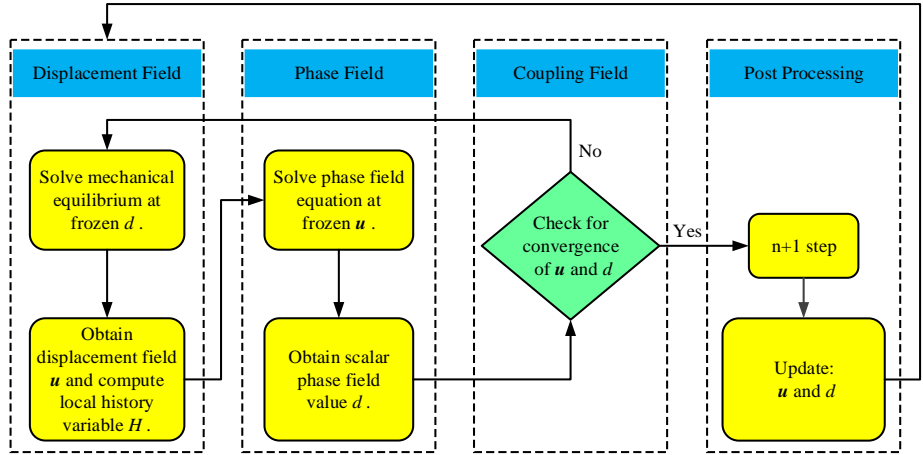

Fig. 3. Flowchart of the staggered scheme

## 3.4. Implementation in AsFem

AsFem is a new open source finite element package, which can be used for phase field modeling and multiphysics simulation. It is written based on C++ and relies on the PETSc [41] library to achieve efficient simulation and simulation. The computing framework is shown in Fig. 4. At present, there are commonly used mechanical models in AsFem. Based on these models, the mechanical problems can be implemented, such as solid mechanics analysis, Cahn Hilliard diffusion simulation, phase field fracture model, linear elastic materials analysis. At the same time, AsFem also has excellent secondary development functions. Users can define their own material model by using the user defined material (UMAT), and their own mechanical model (governing equations) by using the user defined element (UEL). Furthermore, parallel computing is supported in AsFem to improve computing efficiency.

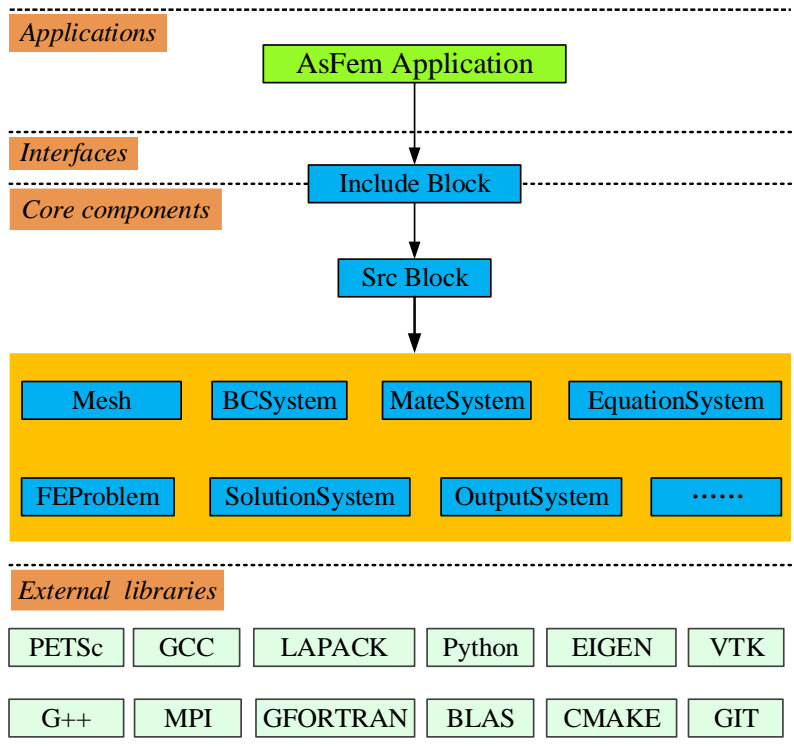

Fig. 4. The frame structure of AsFem

At present, in the open source finite element package AsFem, the phase field model can be easily implemented. The AT2 model is now implemented in AsFem. Further, using UEL and UMAT development, other phase field models can be realized, such as AT1[42], PFM-CZM[43] model. The implementation process in AsFem is shown in Fig. 5.

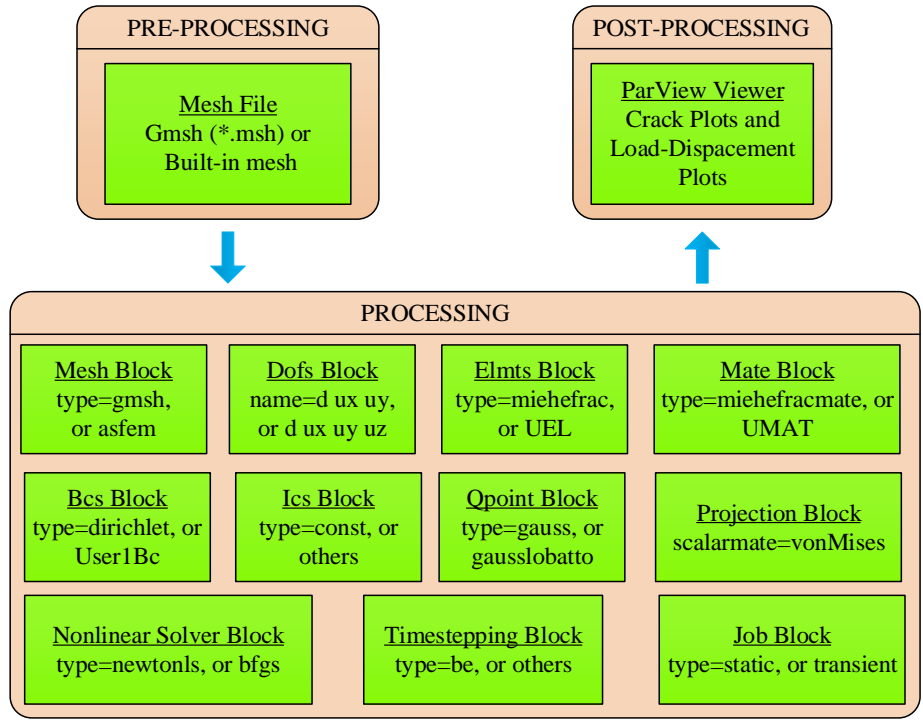

Fig. 5. Phase field model calculation flow in AsFem

# 4. Representative numerical examples

In this section, some examples are presented to show the simulation effect of phase field fracture model in AsFem implementation. These fracture problems can be used to reveal various properties of cracks.

## 4.1. One element test

The first example is a single element model for theoretical comparison. The boundary conditions and the geometry are shown in the Fig. 6. The material properties of the specimen are $E$=210kN/mm$^2$, $\upsilon$=0.3, $G_c$=2.7×10$^{-3}$kN/mm and the length parameter is $l_c$=0.01mm. The phase field parameters and stresses computed by analytical, monolithic, and staggered schemes are shown in the Fig. 7.

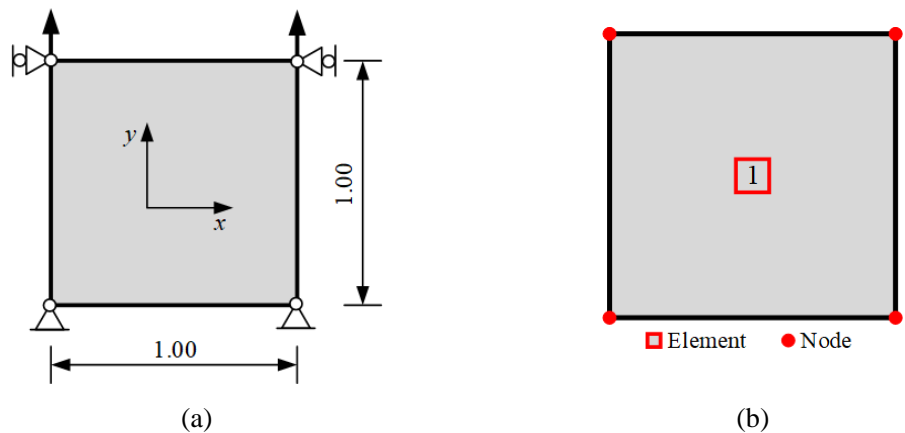

(a)                  (b)

Fig. 6. Geometry and boundary conditions for one element problem: (a) domain and boundary

conditions. (b) discretized domain with one quadrilateral element.

The problem can be solved analytically. If a simple deformation scheme is assumed: $\varepsilon_y \neq 0$, $\varepsilon_x = \tau_{xy} = 0$, the analytical solution of axial stress, elastic strain energy, and the phase field parameter can be calculated as:

$$\sigma_{y,0} = E_{22}\varepsilon_y = \frac{E(1-\nu)\varepsilon_y}{(1+\nu)(1-2\nu)}, \tag{25}$$

$$\psi_0 = \frac{1}{2}\frac{E(1-\nu)}{(1+\nu)(1-2\nu)}\varepsilon_y^2 \tag{26}$$

$$d = \frac{2\psi_0}{\frac{G_c}{l}+2\psi_0} = \frac{\frac{E(1-\nu)}{(1+\nu)(1-2\nu)}\varepsilon_y^2}{\frac{G_c}{l}+\frac{E(1-\nu)}{(1+\nu)(1-2\nu)}\varepsilon_y^2} \tag{27}$$

$$\sigma_y = (1-d)^2 \sigma_{y,0}, \tag{28}$$

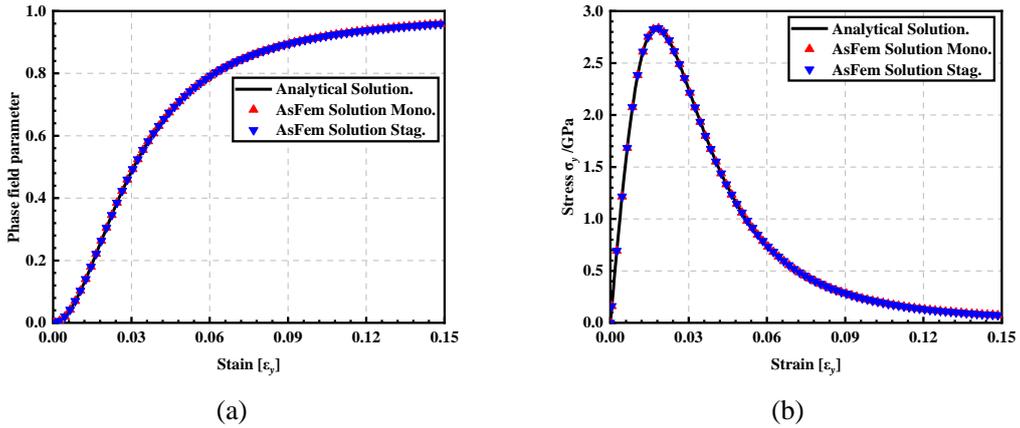

(a) (b)

Fig. 7. One element problem: (a) phase field parameter of axial strain. (b) axial stress of applied axial strain.

Fig. 7(a) shows the phase field computed by analytical, monolithic, staggered scheme, phase field parameter is evolving from zero with the axial strain and reaches a maximum value. Fig. 7(b) shows the evolution of axial stress as a function of applied axial strain. It can be found that the results obtained by different methods are in a good agreement, indicating the feasibility of numerical calculation in AsFem. It can also be observed that the load-bearing capacity of the material degrades with the evolution of the phase field.

*4.2. Single edge notched test*

The second benchmark test is the well known single edge notched tensile and shear sample. Firstly, the specimen is subjected to displacement at the top in $y$ direction to simulate a mode I fracture. The geometry is shown in Fig. 8(a). Then the same specimen is loaded in shear mode, the geometry is shown in Fig. 8(b). The stiffness is $E$=210kN/mm$^2$, $\upsilon$=0.3. The fracture properties are taken identical to Ref. [30] for direct comparison: $l_c$=0.015mm, $G_c$=2.7×10$^{-3}$kN/mm.

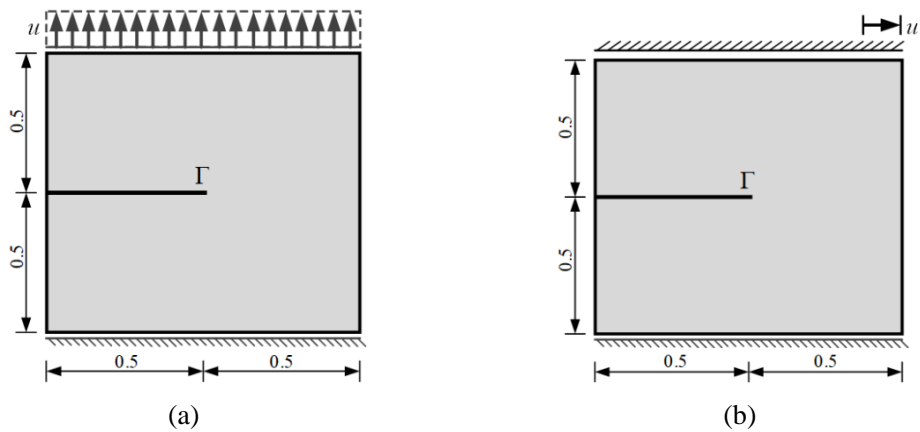

(a)            (b)

Fig. 8. Geometry and boundary conditions of single edge notched specimen: (a) Tensile test. (b) Shear test.

In the tensile test, the monolithic scheme and the staggered scheme are used. For the tensile loading of monolithic scheme, an adaptive step size is applied, which varies from 10$^{-4}$mm to 10$^{-12}$mm. In the staggered scheme, a fixed step size is set, which is 5×10$^{-6}$mm. In shear tests, only the monolithic scheme is adopted, and the shear loading step is applied with an adaptive step, varying from 10$^{-4}$mm to 10$^{-12}$mm. The fracture pattern and load displacement curve of the tensile test are shown in Fig. 9-10, respectively. The fracture pattern and load displacement curve of the shear test are shown in 11-12, respectively.

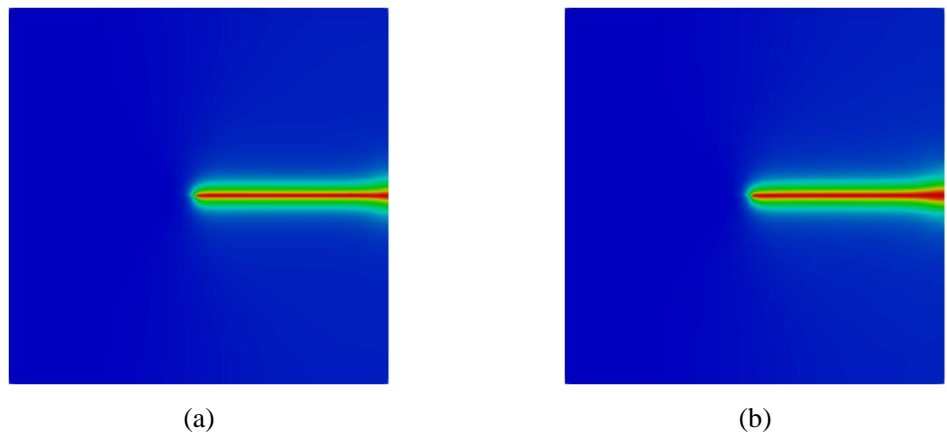

(a)            (b)

Fig. 9. Fracture pattern of single edge notched specimen: (a) monolithic scheme. (b) staggered scheme.

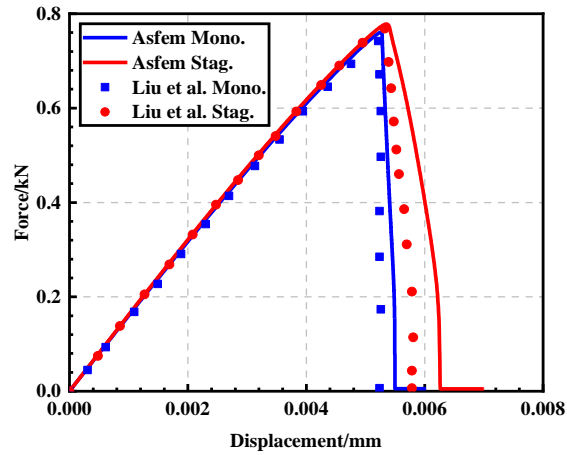

Fig. 10. Load displacement curve of the tensile test.

The fracture pattern for tensile case obtained by two schemes in Fig. 9 is in good agreement with both works of Miehe et al. [11] and Liu et al. [30]. In Fig. 10 the *y* directional reaction force is shown for the tensile test for different schemes. It can be seen that the maximum reaction force value is in agreement, only a small deviation in the propagation period can be observed, the crack propagation calculated by our solution is slowed down ever further.

Further, it can be seen from Fig. 10 that in the monolithic scheme, the force decreases rapidly when the crack starts to propagate, while in staggered scheme, the force decreases slowly when the crack starts to grow. Therefore, the load displacement curve of the monolithic scheme is steeper than that of the staggered scheme.

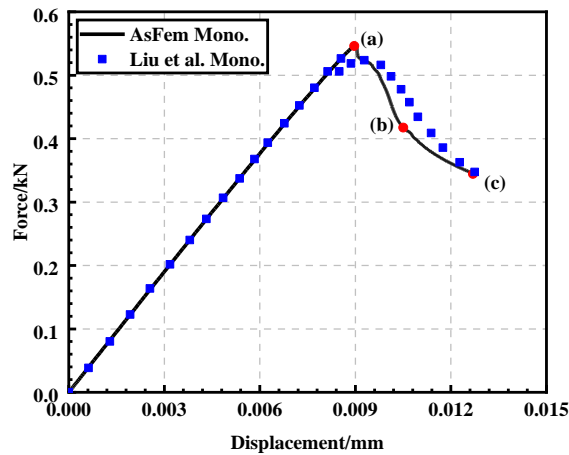

Fig. 11. Load-displacement curve of the shear test

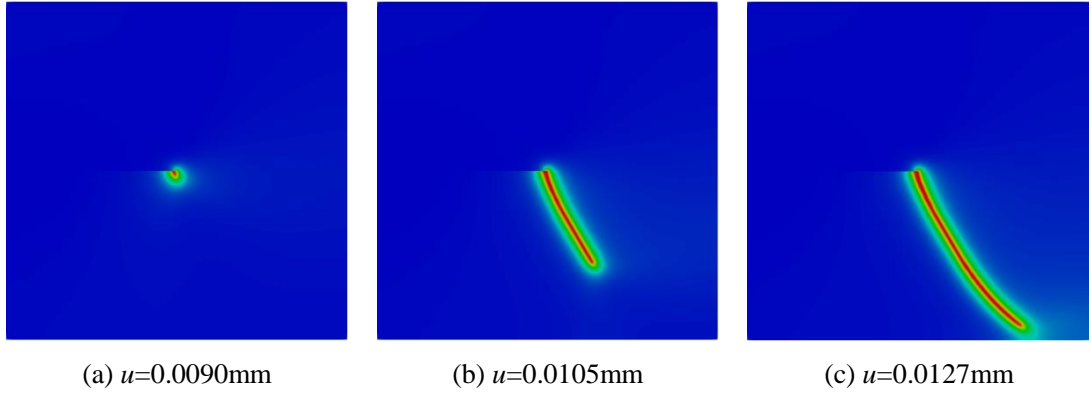

|(a) $u$=0.0090mm | (b) $u$=0.0105mm | (c) $u$=0.0127mm|

Fig. 12. Fracture pattern of shear test for different load steps.

In Fig. 11 the reaction force is shown for the shear test for monolithic schemes. It can be seen that the maximum reaction force value is approximately consistent, only a small deviation in the propagation period can be observed. When the crack starts to expand, there is a sudden drop in the load, and then the nonlinear change begins. Similarly, as it is shown in tensile test, the crack propagation calculated by our solution is slowed down ever further. The fracture pattern for shear case obtained by monolithic scheme in Fig. 12 is in good agreement with the work of Miehe et al. [6,11] and Liu et al. [30].

### *4.3. Asymmetric double notched tensile test*

Now, an asymmetric double notch specimen is used to study multiple crack growth. The geometry is depicted in Fig. 13(a). The following material properties are chosen as: $E$=210kN/mm$^2$, $v$=0.3. $l_c$=0.2mm, $G_c$=2.7×10$^{-3}$kN/mm. The monolithic scheme is adopted, and the loading step is applied with an adaptive step, varying from 10$^{-4}$mm to 10$^{-12}$mm.

In Fig. 13(b) the reaction force is shown for the asymmetric double notched tensile test for monolithic schemes. The reaction force shows similar behavior as we observed in example 4.2 of the tensile test. However, taking a closer look at the two curves, it is found that the force curve of the asymmetric double notched tensile test has turned. This is caused by the interaction between the two cracks in the double notched test and the eventual bending, as shown in Fig. 14(c) and (d). The appearing fracture pattern of the tensile test is shown in Fig. 14. The eventual fracture pattern obtained by phase field method is in good agreement with the literature [31,44]. This example shows the feasibility of simulating multiple crack growth in AsFem.

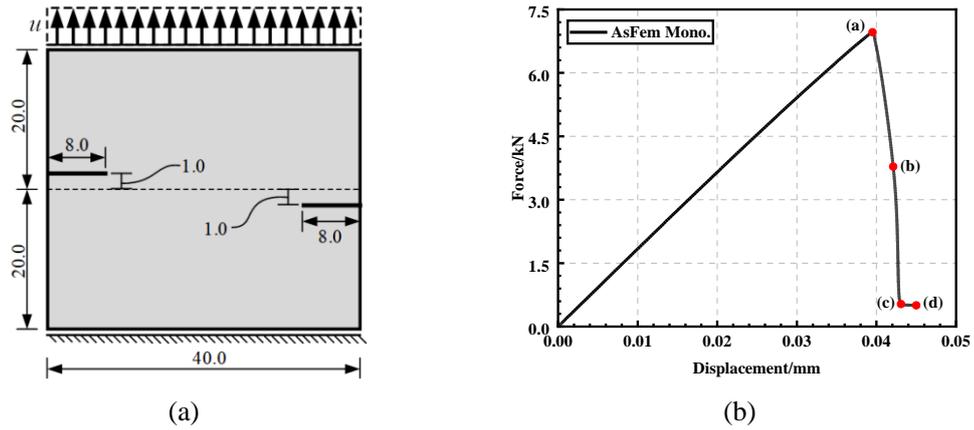

(a)                                  (b)

Fig. 13. Asymmetric double notched tensile test: (a) Geometry and boundary conditions. (b) Load-displacement curve.

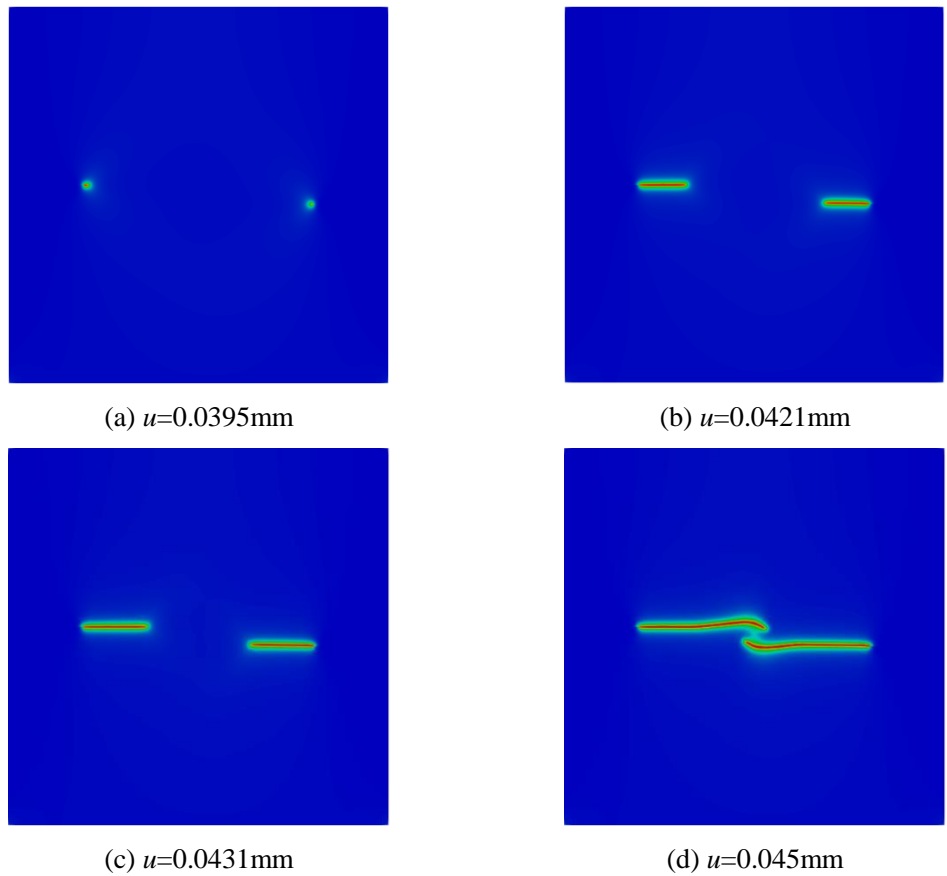

(a) $u$=0.0395mm

(b) $u$=0.0421mm

(c) $u$=0.0431mm

(d) $u$=0.045mm

Fig. 14. Fracture pattern of asymmetric double notched tensile test for different load steps.

## 4.4. Asymmetric notched three point test

Here, an asymmetric three-point bending test is used to study the complex crack trajectory. The experiment and numerical analysis were carried out in the literature [46], and the numerical research was also carried out in the literature [11,31]. The geometry is shown in Fig. 15. The material parameters are $E$=20.8kN/mm$^2$, $v$=0.3, $l_c$=0.02mm, $G_c=1\times10^{-3}$kN/mm. The monolithic scheme is adopted in numerical analysis, and the loading step is applied with an adaptive step, varying from $10^{-4}$mm to $10^{-12}$mm.

Fig. 16 shows the comparison of the fracture pattern achieved by the phase field method with the experimental results and the previous numerical results. In Fig. 16(a), it is the simulation result of phase field method. Fig. 16(b) visualizes the experimental result of Bittencourt et al. [46]. Fig. 16(c) and (d) show the crack results of the model simulated by Miehe et al. [11] and Molnár et al. [31], respectively. The comparison between the results obtained in this paper and the experimental results shows that the phase field method can capture the curvilinear crack mode well.

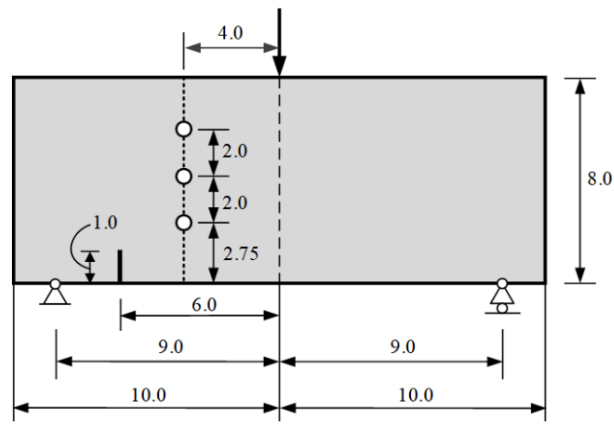

Fig. 15. Geometry and boundary conditions for the asymmetric three point bending test.

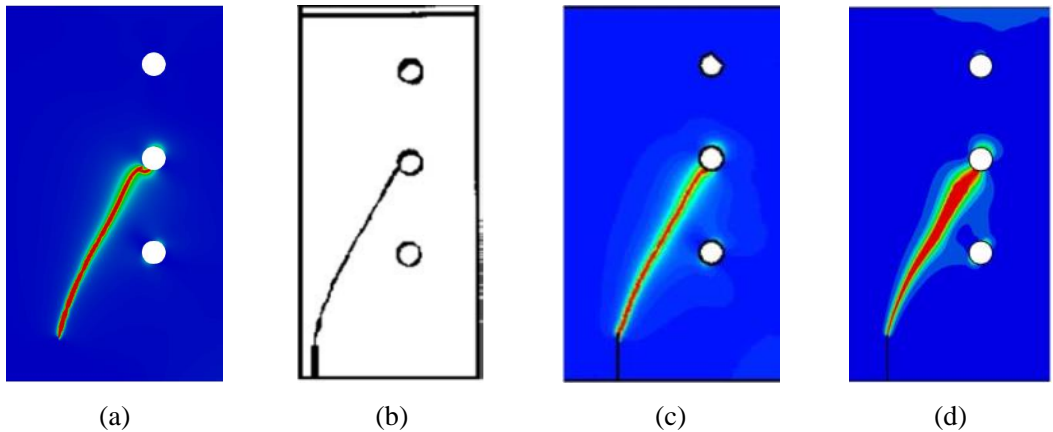

(a)      (b)      (c)      (d)

Fig. 16. Fracture pattern of asymmetric three point bending test: (a) The simulation results of this paper. (b) Experimentally obtained crack patterns by Bittencourt et al. [46]. (c)Crack trajectories of Miehe et al. [11]. (d) Crack trajectories of Molnár et al. [31].

*4.5. Notched bi-material tensile test*

The fracture phenomena of bi-material model under tensile conditions as reported in the previous work of Molnár et al. [31] and Li et al. [45]. The aim here is to demonstrate crack initiation and branching in AsFem. The geometry of the specimen is depicted in Fig. 17(a). The upper material is stiffer and the fracture toughness is higher than the lower material as shown in the Fig. 17(a). The upper material properties are used: $E$=377kN/mm$^2$, $v$=0.3. $l_c$=0.3mm, $G_c$=0.01kN/mm. While the lower material is $E$=37.7kN/mm$^2$, $v$=0.3, $l_c$=0.3mm, $G_c$=0.001kN/mm. There is a notch of 2mm on the lower material. In this test, the monolithic scheme is adopted, and the loading step is applied with an adaptive step, varying from $10^{-4}$mm to $10^{-12}$mm.

The reaction force is shown in Fig. 17(b). It can be observed that the load curve decreases for a short time and then starts to rise again, and the temporary decrease of reaction force does not affect the overall response significantly, namely, it is still increasing until the stiffer material part occurs crack with the continuous loading. Fig. 18 shows the evolution of the damage in the specimen. After initiation the fracture propagates in the lower material part until reaching the material transition zone. When the crack starts to expand in the soft material shown in Fig. 18(a), the force drop begins to occur, corresponding to point (a) in Fig. 17(b), it shows the brittle fracture characteristics of phase field simulation. Then, since crack branching requires less energy than continuing into a harder material, the crack propagates along the interface, as shown in Fig. 18(b). When the energy reaches the energy required for the hard material to expand, it stops expanding at the interface and forms new cracks in the hard material instead, as shown in Fig. 18(c).

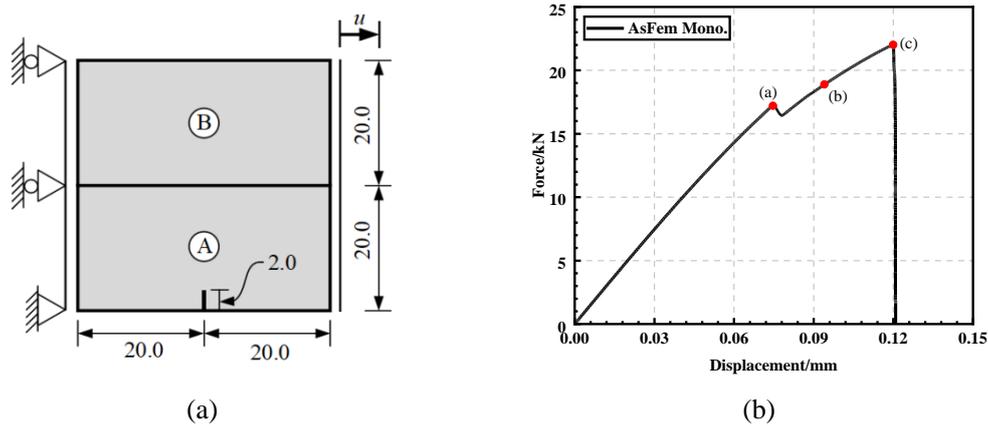

Fig. 17. Notched bimaterial tensile test: (a) Geometry and boundary conditions. (b) Load-displacement curve.

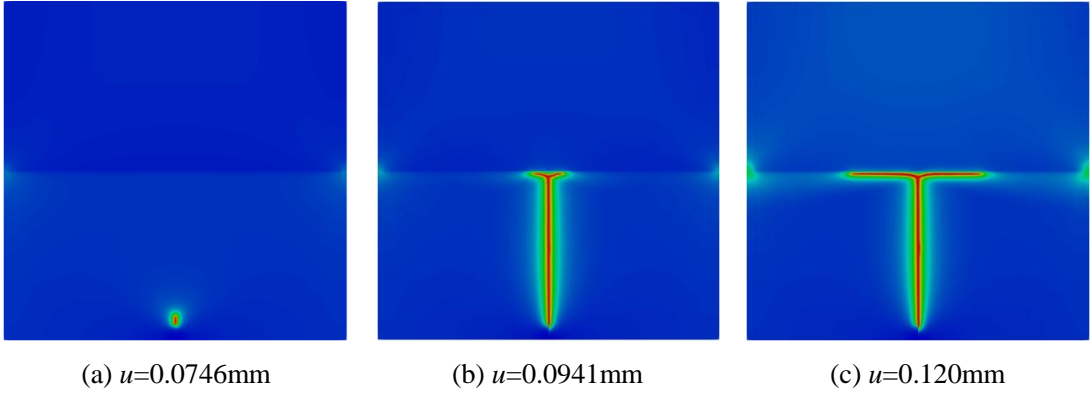

(a) *u*=0.0746mm  (b) *u*=0.0941mm  (c) *u*=0.120mm

Fig. 18. Fracture pattern of notched bimaterical tensile test for different load steps.

### 4.6. Three dimensional single notched plate

To demonstrate the simplicity of the extension from the two dimensional model to three dimensional model in AsFem, we consider a single edge notched specimen to simulate mode I fracture. The geometry and the boundary conditions are shown in Fig. 19(a). The material parameters are $E$=20.8kN/mm$^2$, $v$=0.3, $l_c$=0.2mm, $G_c$=5×10$^{-4}$kN/mm. The monolithic scheme is adopted, and the loading step is applied with an adaptive step, varying from 2.5×10$^{-3}$mm to 10$^{-12}$mm.

The load displacement curve of the three dimensional tensile test are shown in Fig. 19(b). The reaction force exhibits similar behavior to what we observed in example 4.2 of the two dimensional tensile test. The fracture pattern of the three dimensional single notched plate tensile test is shown in Fig. 20. The isosurface of phase field parameter (*d*) is shown with the value larger than 0.95 for the visualization of crack propagation. The result is in agreement with the monolithic solution of Miehe et al. [11] and staggered solution of Molnár et al. [31].

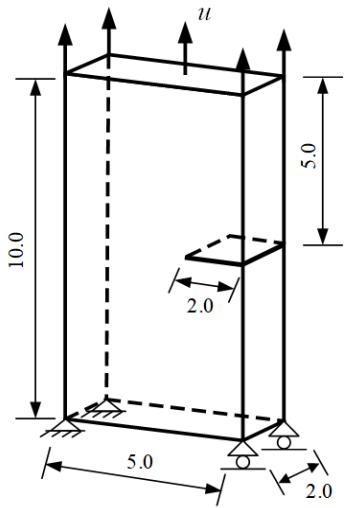
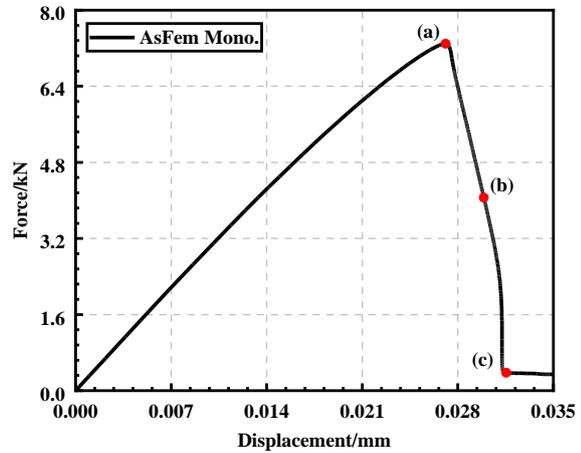

|         (a)          |          (b)          |

Fig. 19. Three dimensional single edge notched tensile test: (a) Geometry and boundary conditions. (b) Load-displacement curve.

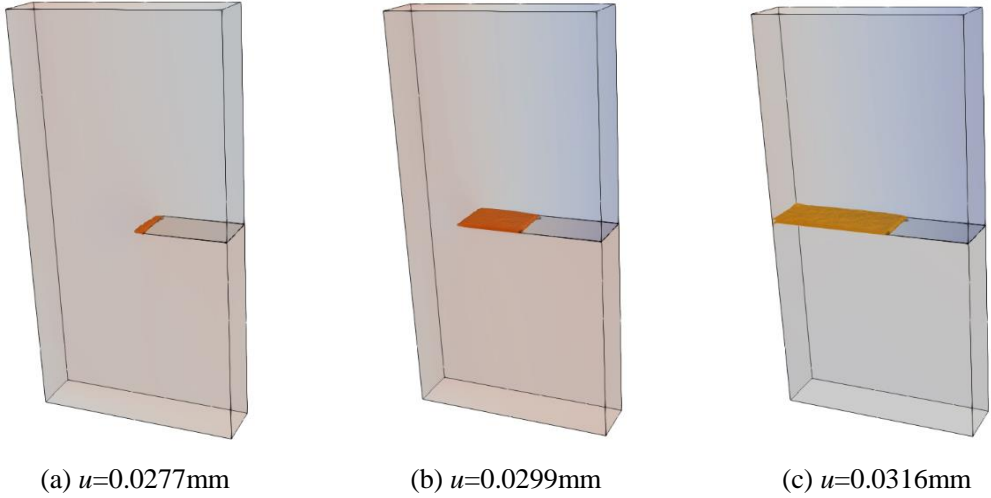

| (a) $u$=0.0277mm | (b) $u$=0.0299mm | (c) $u$=0.0316mm |

Fig. 20. Fracture pattern of three dimensional single edge notched tensile test for different load steps.

## 5. Conclusion

In this work, the phase field methods for brittle fracture problems are implemented in the open source finite element software package AsFem. The simulation results have been verified by the numerical examples in the literature. The displacement and phase field sub-problems are either solved simultaneously or independently decoupled. It is only necessary to establish the connection by using historical variables of elastic strain energy.

The monolithic and staggered scheme are adopted, and the results are compared with those in the literature. It is found that the crack growth in AsFem was consistent with the results in the paper, but the crack growth speed is slightly slower after the crack started. The source code is available as supplemental material with some benchmark examples in this paper. AsFem is a new finite element software package, which enables researchers to simulate not only crack initiation, propagation, curve path, but also other mechanical problems. In the future, AsFem will be more helpful to realize other types of multi-physical phase field models[47,48].

## Acknowledgment

Y. Yu would like to thank Dr. Bai of Max-Planck-Institut für Eisenforschung for discussions on AsFem implementation in this work.